\documentstyle[aps,twocolumn,epsf,tighten]{revtex}

\begin{document}
\pagestyle{empty}                                      %%%To be commented
\draft
\vfill
%
%%
%\twocolumn[\hsize\textwidth\columnwidth\hsize\csname%
%@twocolumnfalse\endcsname]%
%%
\title{The Glueball Spectrum from a Potential Model}
\vfill
\author{$^{1}$Wei-Shu Hou and $^{2}$Gwo-Guang Wong}
\address{
\rm $^{1}$Department of Physics, National Taiwan University,
Taipei, Taiwan 10764, ROC\\
and\\
\rm $^{2}$Department of International Trade, Lan Yang Institute of Technology,
Toucheng, Ilan, Taiwan 26141, ROC
}
\date{\today}

\vfill
\maketitle
\begin{abstract}

The spectrum of two-gluon glueballs below 3 GeV is investigated in
a potential model with dynamical gluon mass using variational
method. The short distance potential is approximated by one-gluon
exchange, while the long distance part is taken as a breakable
string. The mass and size of the radial as well as orbital
excitations up to principle quantum number $n=3$ are evaluated.
The predicted mass ratios are compared with experimental and
lattice results.
\end{abstract}
\pacs{PACS numbers: 12.39.MK}

%\vskip2pc
%
\pagestyle{plain}

%\section{Introduction}

Quantum Chromodynamics (QCD) is widely accepted as the theory of
strong interactions. It is generally believed that the gluon
self-coupling in QCD implies the existence of bound states of
confined gluons known as glueballs. The experimental discovery of
these glueballs would be very important and would give further
support to the theory of QCD. However, numerous technical
difficulties have so far hampered our unequivocal identification
of glueballs by experiment, largely because glueball states can
mix strongly with nearby $q \bar q$ resonances. Nevertheless, the
estimation of mass and size of pure gluon glueball states should
still be pursued. This could guide experimental searches, as well
as provide calibration for models of glueballs.

Over the past 20 years there has been an on-going effort to obtain
a nonperturbative form for the gluon propagator.
Perhaps one of the most interesting result is that
the gluon may have a dynamically generated mass~\cite{cornwall}.
The existence of a mass scale, or the absence of a pole at $k^2=0$,
is natural if one assumes that gluons do not propagate to infinity,
{\it i.e.}, these propagators describe confined gluons.
The concept of massive gluon has been widely used in
independent field theoretic studies, and
examples about the consequences of massive gluons can be found
in the literature~\cite{parisi,halzen,field,Consoli,Papavassiliou,Natale}.

In this paper, we focus on the calculation of two-gluon glueball
systems and extend our previous work~\cite{hounwong} on the
estimation of the mass and size of low lying glueball states,
using the variational method in the potential model of Cornwall
and Soni~\cite{cands,Hou}. The main feature of the present work is
the consideration of radial as well as orbital excitations, up to
principle quantum number $n=3$.

To exhibit both asymptotic freedom and the non-Abelian nature of
QCD, gluon dynamics can be described as massive spin-one fields
interacting through one-gluon exchange and a breakable string. At
short distance the effective coupling constant
of the gluon-gluon interaction %$\lambda$
becomes small and the interaction
can be treated perturbatively.
The short distance potential is approximated by one-gluon exchange and
can be extracted from the tree-level Feynman amplitude of Fig. 1,
\begin{equation}
V({\bf r})=\int {d^3q \over (2\pi)^3}
{ie^{i{\bf q}\cdot{\bf r}} \over 4\sqrt{E_{1f}E_{2f}E_{1i}E_{2i}}}
i{\cal M}_{fi},
\label{m34}
\end{equation}
where ${\bf q}$ is the momentum transfer of the system.
At long distance, the non-Abelian nature of QCD implies
gluon confinement via nonperturbative effects.
These nonperturbative effects are implemented by
introducing a string potential $V_{\rm str}$ which is assumed to be spin
independent,
\begin{equation}
V_{\rm str}=2m(1-e^{-\beta mr}),
\end{equation}
where $\beta$ is related to the adjoint string tension $K_A$ via
\begin{equation}
\beta={K_A \over 2m^2}.
\end{equation}
In the potential $V_{\rm str}$, the color screening of gluons is
brought about by a breakable string; that is, the adjoint string
breaks when sufficient energy has been stored in it to materialize
a gluon pair. This form of the string potential simulates the
inter-gluonic potential as seen in lattice
calculations~\cite{Bernard}.

Thus the gluon-gluon potential relevant for two-gluon glueballs is~\cite{hounwong}
\begin{eqnarray}
V_{2g}(r) &=&-\lambda\Biggl\{
\biggl[{1 \over 4}+{1 \over 3}{\bf S}^2
+{3 \over 2m^2}({\bf L}\cdot{\bf S})
{1 \over r}{\partial \over \partial r}\biggr. \Biggr.
 \nonumber \\
&& -{1 \over 2m^2} \biggl. \Bigl(({\bf S}\cdot\nabla)^2
-{1 \over 3}{\bf S}^2\nabla^2\Bigr)\biggr]
{e^{-mr} \over r}  \nonumber \\
&& +\Biggl. \Bigl(1-{5 \over 6}{\bf S}^2\Bigr)
    {\pi \over m^2}\delta^3({\bf r})\Biggr\}
+2m(1-e^{-\beta mr}), \label{m82}
\end{eqnarray}
where $\lambda$ is defined as
\begin{equation}
\lambda \equiv {3g^2 \over 4\pi},
\end{equation}
and is related to the strong coupling strength of the process.
The terms containing $\lambda$ are from the short distance
\begin{figure}[b!]
\centerline{{\epsfxsize3.2 in \epsffile{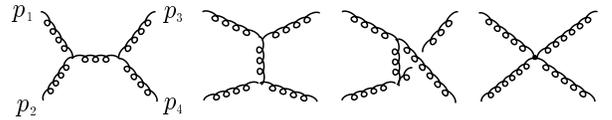}}}
\smallskip
\caption{Diagrams for $gg \rightarrow gg$ scattering.}
\end{figure}
\noindent potential. Note that the color wavefunctions have been
contracted out with the structure constants in gluon
vertices~\cite{hounwong}, so that the strength of the gluon-gluon
coupling is three times as large as that of the gluon-quark
coupling. In Eq. (\ref{m82}), ${\bf S}\equiv{\bf S}_1+{\bf S}_2$
is the total spin of the 2-gluon glueball. Note also that ${\bf
S}_1$ acts on the polarization vectors ${\bf e}_1$ and ${\bf e}_3$
while ${\bf S}_2$ acts on the polarization vectors ${\bf e}_2$ and
${\bf e}_4$. Each spin operator ${\bf S}_n=(S^1,S^2,S^3)$ is
defined as
%
%\[
%S^1=\pmatrix{0 & 0 & 0\cr
%0 & 0 & -i\cr
%0 & i & 0\cr},\
%S^2=\pmatrix{0 & 0 & i\cr
%0 & 0 & 0\cr
%-i & 0 & 0\cr},\\
%S^3=\pmatrix{0 & -i & 0\cr
%i & 0 & 0\cr
%0 & 0 & 0\cr},
%\]
%or
%\begin{equation}
$(S^k)_{ij}=-i\epsilon^{ijk}$,
%\label{m38}
%\end{equation}
which satisfy
\begin{equation}
[S^i,S^j]=i\epsilon^{ijk}S^k,
\label{m39}
\end{equation}
{\it i.e.} $S^1$, $S^2$ and $S^3$ are
${\rm SU}(2)$ group generators as desired.

For this gluon-gluon potential we are left with three parameters:
effective gluon mass $m$, string breaking parameter $\beta$ and
adjoint strong coupling constant $\lambda$. In this model the
constituent gluon mass is evaluated to be 600--700 MeV, which is
roughly twice the constituent quark mass~\cite{hounwong}. In the
intermediate or at least at long distances, the fundamental string
tension $K_F$ is one of the most fundamental physical quantities
in quark confinement, and is related to the Regge slope $\alpha'$
by~\cite{GGBT}
\begin{equation}
K_F=\frac{1}{2\pi \alpha'}=0.18\ {\rm GeV^2},
\end{equation}
with the experimental value for $\alpha'\simeq $0.9 GeV$^{-2}$.
The adjoint string tension $K_A$ for gluon confinement can be related by the
strong evidence of Casimir scaling hypothesis~\cite{Bali3} on the lattice via
\begin{equation}
\frac{K_A}{K_F}\approx \frac{9}{4}.
\end{equation}
Thus, the string breaking parameter $\beta$ is about 0.4--0.6. For
the gluon propagator with a dynamical mass the adjoint strong
coupling constant $\lambda$ at one-loop level turns out to
be~\cite{Natale}
\begin{equation}
\lambda (Q^2)=\frac{36 \pi}{(33-2 n_f) \rm{ln}[Q^2+\xi m^2]/\Lambda^2]},
\end{equation}
where $\xi \approx 4$. The higher order correction at two loop
level does affect the value, as shown in the case of the
fundamental strong coupling constant $\alpha_{\rm
s}(Q^2)$~\cite{CKS}. However, it has not been solved so far in the
literature. Nevertheless, $\lambda $ is expected to be in the
range of $1.5\pm 0.5$.

The invariance of charge conjugation and parity for strong
interactions implies that a two-gluon glueball must have quantum
numbers $J^{PC}=J^{(-1)^{L}(+)}$. On the other hand, since the
gluon is a spin 1, color octet boson, the color singlet
wavefunction of a 2-gluon glueball is symmetric. Hence, a
symmetric spin wavefunction with spin 0 or 2 must be accompanied
by a symmetric spatial wave function with an even value of orbital
angular momentum, and an antisymmetric spin wavefunction with spin
1 must be accompanied by an antisymmetric spatial wave function
with an odd value of orbital angular momentum. In this way one can
count the number of states. For example, we have two S-wave, three
P-wave and six D-wave states with definite $J^{PC}$ corresponding
to principle quantum number $n=3$.

The Hamiltonian of two-gluon glueball system is
\begin{equation}
H=2m-{1\over m}\nabla^2+V_{2g}.
\end{equation}
Since the color wavefunction has been contracted out, the remaining
wavefunction is of the form
\begin{eqnarray}
& &\Psi_{nljm}({\vec r})\equiv \vert nljm \rangle \propto \psi_{nl}(r) \vert jm
\rangle \nonumber \\
&\equiv&\psi_{nl}(r)\sum_{m=m_{l}+m_{s}} \langle lm_lsm_s \vert jm \rangle
                                               Y_{lm_l}(\theta,\
                                               \phi)\chi_{sm_s},
\label{m93}
\end{eqnarray}
where
\begin{equation}
\chi_{sm_s}=
%&\equiv&\psi_{nl}(r)\sum_{m_{l}m_{s}}\Bigl\{Y_{lm_l}
           \sum_{\lambda_1\lambda_2} \langle1\lambda_11\lambda_2 \vert sm_s \rangle
                                  {\bf e}_1^{(\lambda_1)}{\bf e}_2^{(\lambda_2)}.
\label{m193}
\end{equation}
The trial radial wavefuctions $\psi_{nl}(r)$ are constructed by
orthogonality as follows,
\begin{eqnarray}
\psi_{10}(r)&\propto& e^{-a^2 m^2 r^2}, \nonumber \\
\psi_{20}(r)&\propto& (1-a^2 m^2 r^2)e^{-a^2 m^2 r^2/2}, \nonumber \\
\psi_{21}(r)&\propto& a^2 m^2 r^2e^{-a^2 m^2 r^2/2}, \nonumber \\
\psi_{30}(r)&\propto& \Bigl ( 1-\frac{62}{51}a^2 m^2 r^2+\frac{13}{68}a^4 m^4 r^4
\Bigr )
                                        e^{-a^2 m^2 r^2/4}, \nonumber \\
\psi_{31}(r)&\propto& a^2 m^2 r^2 \Bigl (1-\frac{3}{14}a^2 m^2 r^2 \Bigr )
                                            e^{-a^2 m^2 r^2/4}, \nonumber \\
\psi_{32}(r)&\propto& a^4 m^4 r^4 e^{-a^2 m^2 r^2/4},
\label{m94}
\end{eqnarray}
where $a$ is the variational parameter. The spin wavefunction
$\chi_{sm_s}$ is constructed by the direct product of two gluon
polarization vectors ${\bf e}_1^{(\lambda_1)}$ and ${\bf
e}_2^{(\lambda_2)}$ and, in turn, the total angular momentum
eigenstate $\vert jm \rangle$ can be constructed by the direct
product of orbital eigenstate, the spherical harmonics
$Y_{l{m_l}}(\theta, \ \phi)$, and the spin wavefunction
$\chi_{sm_s}$. The coefficients $\langle 1\lambda_11\lambda_2
\vert sm_s \rangle$ and $\langle lm_lsm_s \vert jm \rangle$ in
Eqs. (\ref{m93}) and (\ref{m193}) are just the Clebsch-Gordan
coefficients.

When considering the case with $L=0$~\cite{hounwong}, we drop the
spin-orbit and tensor terms in Eqs. (\ref{m82}). However, they do
contribute for the case with $L>0$. Since only the tensor term in
Eq. (\ref{m82}) is related to the total angular momentum
eigenstates $\vert jm \rangle$, we need to calculate their
expectation values. Defining the tensor operator as,
\begin{equation}
{\bf T}={\bf S}^2-3({\bf S}\cdot {\bf {\hat r}})^2,
\end{equation}
integrating out the spherical harmonics and doing spin algebraic
calculation on spin wavefunctions, we obtain
\begin{equation}
\langle 0\ 0 \vert {\bf T} \vert 0\ 0 \rangle =0, \ \ \ \ \ \ {\rm for}\ l=0,\ s=0,
\nonumber\\
\end{equation}
and
\begin{equation}
\langle 2\ m \vert {\bf T} \vert 2\ m \rangle =0, \ \ \ \ \ \ {\rm for}\ l=0,\ s=2.
\end{equation}
For $l=1$ and $s=1$ we have
\begin{eqnarray}
\langle 2m \vert {\bf T} \vert 2m \rangle &=&\frac{1}{5}, \nonumber \\
\langle 1m \vert {\bf T} \vert 1m \rangle &=&-1,\nonumber \\
\langle 0\ 0 \vert {\bf T} \vert 0\ 0 \rangle &=&2.
\end{eqnarray}
For $l=2$ and $s=0$, we have 
\begin{equation}
\langle 2\ m \vert {\bf T} \vert 2\ m \rangle =0,
\end{equation}
and for $l=2$ and $s=2$, we have
\begin{eqnarray}
\langle 4m \vert {\bf T} \vert 4m \rangle &=&\frac{12}{7}, \nonumber \\
\langle 3m \vert {\bf T} \vert 3m \rangle &=&-\frac{24}{7}, \nonumber \\
\langle 2m \vert {\bf T} \vert 2m \rangle &=&-\frac{9}{7}, \nonumber \\
\langle 1m \vert {\bf T} \vert 1m \rangle &=&3,\nonumber \\
\langle 0\ 0 \vert {\bf T} \vert 0\ 0 \rangle &=&6.
\end{eqnarray}
Hence the contribution to the glueball mass $M$ can be written as
\begin{equation}
%\frac
{M}/{m}=2+E_{\rm K}+E_{\rm Y}+E_{\delta}+E_{LS}+E_{\rm {T}}+E_{\rm
str},
\end{equation}
where
\begin{eqnarray}
E_{\rm K}&=&-\frac{1}{m^2}{\left < nl|\nabla^2|nl \right >}, \nonumber \\
E_{\rm Y}&=&-\frac{\lambda}{m}\left [\frac{1}{4}+\frac{1}{3}S(S+1)\right ]
                       \left < nl  \Bigg \vert \frac{e^{-mr}}{r}  \Bigg \vert nl
\right > , \nonumber \\
E_{\delta}&=&-\frac{\lambda \pi}{m^3}\left [1-\frac{5}{6}S(S+1)\right ]
                                           \left < nl \vert \delta^3({\bf r})\vert
nl \right >, \nonumber \\
E_{LS}&=&\frac{3\lambda}{2m^3}{\bf L}\cdot{\bf S}
                                           {\left <  nl \Bigg \vert \left
(\frac{1}{r^2}+\frac{m}{r}\right )
                                                             \frac{e^{-mr}}{r}
\Bigg \vert nl \right >}, \nonumber \\
E_{\rm T}&=&-\frac{\lambda}{2m^3}{\left < jm|{\bf T}|jm\right >} \nonumber \\
&& \ \ \ \ \ \
                                           {\left < nl \Bigg | \left
(\frac{1}{r^2}+\frac{m}{r}+\frac{m^2}{3}\right )
                                                             \frac{e^{-mr}}{r}
\Bigg | nl \right >}, \nonumber \\
E_{\rm str}&=&2{\left < nl \vert \left (1-e^{-\beta mr}\right ) \vert nl \right >},
\end{eqnarray}
and $|nl\rangle \equiv \psi_{nl}(r)$.

We then use the variational method with trial radial wavefunction
$\psi_{nl}(r)$ to evaluate the glueball mass $M$ and
root-mean-square radius $r_{\rm {rms}}=\sqrt{\langle r^2 \rangle}$
for each glueball state. The lightest scalar and tensor glueballs
have been investigated in Ref.~\cite{hounwong}. For the lightest
scalar glueball there is an attractive $\delta$-function term in
the potential and hence the Hamiltonian is unbounded from below.
It has been conjectured that this maximum attraction channel in
$0^{++}$ state could be related to the gluon condensation that
triggers confinement. We proposed a physical solution by smearing
the gluon fields; that is, we replace the $\delta$ function by the
smearing function
\begin{equation}
D(r)={{k^3m^3}\over \pi^{3\over 2}}e^{-k^2m^2r^2},
\label{m38}
\end{equation}
which approaches $\delta^3({\bf r})$ for $k\longrightarrow\infty$.
In contrast to the lightest scalar glueball, the lightest tensor
glueball is stable since the $\delta$-function term is repulsive.
As the variation is slight in the mass and size
estimation~\cite{hounwong} of a glueball in the $\lambda$--$\beta$
parameter space, we take the central values of $\lambda =$1.5, and
$\beta =$0.5~\cite{foot}.

The lattice result of $M(1^1S_0),$ $M(1^5S_2)=$ 1730, 2400 MeV is
taken as input for the lightest $0^{++}$ and $2^{++}$ glueballs,
respectively~\cite{Morningstar,Bugg}. From the $2^{++}$ input, we find
the constituent gluon mass $\sim $ 670 MeV, about twice
the constituent quark mass, and the lightest tensor glueball is found to
have the typical hadron size of $\sim $ 0.8 fm.

The value for $k$ in Eq.~(\ref{m38}) is fixed by the mass ratio
\begin{equation}
{M(1^5S_2) \over M(1^1S_0)}\cong 1.39, \label{rat}
\end{equation}
from the converging lattice and experimental results. The size of
the lightest scalar glueball is found to be a mere $\sim$ 0.1 fm.
We note that, although we always have an attractive
$\delta$-function term in the potential for scalars, only for the
lightest scalar glueball is the smearing of gluon fields needed.
For all other scalars, the $\delta$-function potential gives very
small mass corrections. One can consider the sum of kinetic energy
and $\delta$-function terms as the effective kinetic energy.
We find that the sum contributes less than 12$\%$ to all glueball
masses, except for the lightest scalar glueball, which is 43$\%$.
This may be marginal for a nonrelativistic treatment, and
stretching the applicability of our relativistic expansion.

We list the mass spectrum and size of two-gluon glueballs up to
$n=3$ in Table 1.

\begin{table}[b!]
{Table 1.  Masses and sizes of two-gluon glueballs up to
$n=3$, with the lightest $0^{++}$ and $2^{++}$ masses taken as input.}%\\[1ex]
  \begin{tabular}{c|c|c|c|c|c}
%  {p{0.6cm}|p{0.6cm}|p{0.6cm}|p{2.0cm}|p{2.0cm}|p{2.0cm}}
    n  & L & S & $J^{PC}$ & M (MeV) &  $r_{\rm rms}$ (fm)  \\ \hline
                                  &                        &0& $0^{++}$ & \{1730\}
&  0.1              \\
      \raisebox{1ex}{1} & \raisebox{1ex}{0} &2& $2^{++}$ & \{2400\} &   0.8
     \\ \hline
                                  &                             &0& $0^{++}$ & 2710
&  2.0               \\
                                  & \raisebox{1ex}{0}  &2& $2^{++}$ & 2730 &  1.9
     \\ \cline{2-6}
                             2   & &  & $0^{-+}$  &  2570    &  0.7     \\
                                  & 1 &1 & $1^{-+}$  &  2605 &  1.0     \\
                                  & & & $2^{-+}$  &  2615  & 1.1      \\ \hline
                                  &                             &0 & $0^{++}$ &
 2790 & 2.6     \\
                                  &\raisebox{1ex}{0}  &2 & $2^{++}$ &   2810 & 2.4
  \\  \cline{2-6}
                                  &                             &   & $0^{-+}$  &
 2765 & 2.4     \\
                                  &               1            &  1& $ 1^{-+}$ &
 2770 & 2.4     \\
                                  &  & & ${2^{-+}}$  &  2775  & 2.4\\ \cline{2-6}
                     3           &             2              & 0& $2^{++}$ & 2700
& 1.6     \\ \cline{2-6}
                                  &                             &   & $0^{++}$ &
 2685 & 1.1  \\
                                  &                             &   & $1^{++}$ &
 2690 & 1.3 \\
                                  &             2              &  2& $2^{++}$ &
 2693 & 1.5   \\
                                  &                             &   & $3^{++}$ &
 2694  &  1.6  \\
                                  & & &$4^{++}$ &  2695 & 1.7
  \end{tabular}
\end{table}

%Besides the constituent term $2m$, we see that the glueball 5
%mass mainly comes from string energy and then from kinetic energy.
%The contributions from Yukawa, ${\bf L}\cdot {\bf S}$, and tensor forces are
%relatively small for forming a stable state.
%Nevertheless, these terms
%change a lot with the variational parameter $a$. The balance among these
%terms become sensitive to generate a stable glueball state.
%From our calculation four states $2(0^{-+})$, $3(0^{++})$, $3(1^{++})$
%and $3(2^{++})$ have no solutions in the suitable physical region of parameter
%space $(\lambda, \beta)$. However,

From Table 1 we see that the glueball mass $M$ decreases with
increasing orbital angular momentum $L$ for fixed $n$. For fixed
$n$ and $L = 0,\ 1$ (2), $M$ increases (decreases) with 
total spin angular momentum $S$.
In turn, both $M$ and size increase with total angular momentum
$J$ for fixed $n$, $L$, and $S$. On the other hand, $M$ increases
with $n$ at fixed $L$ and $S$. All these increments are slight,
except for the $n = 1$ case where the attractive $\delta$
potential is present.

Our calculations show that glueball masses are almost independent
of $\lambda$, and increases with $\beta$ only slightly
\cite{foot}. Although our pure gluon glueballs are different from
real glueball states which can mix strongly with nearby $q\bar q$
resonances, the better way to compare with experimental data is to
take mass ratios, either to eliminate the $\lambda$ and $\beta$
dependence, or to reduce the effect of mixing. We list in Table 2
the glueball masses in increasing order, together with their
corresponding mass ratios with respect to the lightest tensor
glueball. Comparison with Table 1 shows that, for a given quantum
number, the ordering in mass is also the ordering in size.

\begin{table}[t!]{Table 2.  Comparison of predicted glueball
masses (mass ratios, normalized to lightest $2^{++}$) with
lattice~\cite{Morningstar} and sample experimental data. The
superscripts $a$, $b$, $c$ and $d$ indicate data coming
from~\cite{Liunwu},~\cite{Bugg},~\cite{Anisovich} and~\cite{Zou},
respectively. The experimental numbers are not meant to match the
states listed to the left. The last three entries with $3g$ in
front of the $J^{PC}$ are for three gluon
glueballs~\cite{hounwong}. The lightest $0^{++}$ and $2^{++}$
masses are taken as input. All masses are in MeV units.}%\\[1ex]

\begin{tabular}{cccc}
    $J^{PC}$    & Constituent    & Lattice
                      &    Experiment  \\ \hline
    & \{1730\} (0.72) \         & \ 1730 (0.72) \             & \  $1500^{b}$ (0.76)   \ \\  {$0^{++}$}
    & \ 2685\ \ \ (1.12) \         & \ 2670 (1.11) \             & \  $2105^{b}$ (1.06)   \ \\    %{\raisebox{0.5ex}%

    & \ 2710\ \ \ (1.13) \         &                                     & \  $2320^{c}$ (1.17)   \ \\
    & \ 2790\ \ \ (1.16) \         &                                     &
\ \\ \hline
    & \{2400\} (1.00) \         &  \ 2400 (1.00) \             & \  $1980^{b}$ (1.00)    \ \\
    & \ 2693\ \ \ (1.12) \         & \ 3290 (1.37) \             & \  $2020^{d}$ (1.02)    \ \\
    $2^{++}$& \ 2700\ \ \ (1.13) \         &                       & \  $2240^{d}$ (1.13)    \ \\
    & \ 2730\ \ \ (1.14) \          &                                     & \   $2370^{d}$ (1.20)   \ \\
    & \ 2810\ \ \ (1.17) \          &                                     &                                     \ \\ \hline
    {$0^{-+}$}&  \ 2570\ \ \ (1.07) \        & \ 2590 (1.08) \              & \  $2140^{d}$ (1.08)   \ \\
%{\raisebox{0.5ex}%
    &  \ 2765\ \ \ (1.15) \         &                                     & \ $2190^{b}$ (1.11)   \ \\  \hline
   {$1^{-+}$} &  \ 2605\ \ \ (1.09) \         &                                     &                    \ \\
%{\raisebox{0.5ex}%
    &  \ 2770\ \ \ (1.15) \         &                                     &                                     \ \\ \hline
   {$2^{-+}$} &  \ 2615\ \ \ (1.09) \         & \ 3100 (1.29) \             & \   $2040^{d}$ (1.03)     \ \\    %{\raisebox{0.5ex}%
    &  \ 2775\ \ \ (1.16) \         & \ 3890 (1.62) \             & \ $2300^{d}$ (1.16)        \ \\ \hline
    $1^{++}$& 2690\ \ \ (1.12)\          &                        & \ $2340^{d}$ (1.18)        \ \\  \hline
    {$3^{++}$} & \ 2694\ \ \ (1.12) \         & \ 3690 (1.54) \              & \ $2000^{d}$ (1.01)     \ \\  %{\raisebox{0.5ex}%
    &                                &                                      & \ $2280^{d}$ (1.15)     \ \\ \hline
   {$4^{++}$} & \ 2695\ \ \ (1.12) \         &  $3650^{a}\ (1.52)$\    & \ $2044^{d}$ (1.03)     \ \\  %{\raisebox{0.5ex}%
    &                                &                                      &  \ $2320^{d}$ (1.17)     \ \\ \hline
    $3g(0^{-+})$&3780\ \ \ (1.58)          & \ 3640 (1.52) \             &                     \ \\
    $3g(1^{--})$& 3680\ \ \ (1.53)          & \ 3850 (1.60) \             &                     \ \\
    $3g(3^{--})$& 3690\ \ \ (1.54)          & \ 4130 (1.72) \              &                     \ \\
  \end{tabular}
\end{table}

Although our predictions rely on the inputs of the lightest scalar
and tensor glueball masses, 1730 and 2400 MeV respectively from
lattice calculation, we find that our predicted mass ratios do
find experimental correspondence, sometimes even better than
comparing with lattice calculations. This may be due to the rather
straightforward physical picture of a potential model. However,
the experimental situation is far from settled, and comparison
with lattice is quite necessary.

As stated, the lightest tensor glueball is much more stable within
the model than the lightest scalar glueball, hence all mass ratios
are normalized to this state. We find that, except for the
peculiar lightest $0^{++}$ state, all ratios of two-gluon glueball
masses are of order 1, and mass ratios of three-gluon to two-gluon
glueballs are of order 3/2, respectively. This is consistent with
a constituent picture. We then need to understand the difference
from lattice spectrum, which suggests a considerably higher
excitation energy.

Going back to Table 1, we note that the constituent picture gives
certain multiplet structures governed mostly by $n$ and $L$, with
relatively small splittings in $S$ (except for $n = 1$). We note
that there are spin-orbit and tensor forces at work. Thus, the
``first excitation" from the $n = 1$ states seem to be the $(n,\
L,\ S) = (2,\ 1,\ 1)$ (rather than (2, 0, 0)) states of $0^{-+}$,
$1^{-+}$ and $2^{-+}$. The masses are all around 2600 MeV, or
$\sim 1.08$ times the lightest $2^{++}$. They are also of similar
size as the lightest $2^{++}$, the $0^{-+}$ actually smaller,
hence they are likely more relevant in production processes.

The $0^{-+}$ may be most interesting since it is the lightest
pseudoscalar glueball. Its mass of 2570 MeV is in remarkable
agreement with the lattice result of 2590 MeV. The heaviness
suggests
\cite{Hou:1997jb} that there may be little glueball
mixture in $\eta$ and $\eta^\prime$, and OZI rule violation in
these mesons arise from vacuum effects. The $1^{-+}$ state with
exotic quantum numbers cannot be a $q\bar q$ meson and is also
very interesting. Again, its heaviness compared to hybrid
candidates such as $\hat \rho$(1600)~\cite{Page} suggest that
search for lightest hybrids are not complicated by presence of
glueballs. For the $2^{-+}$, the mass of 2615 MeV in constituent
model is in contrast with the lattice result of 3100 MeV, which
has a 500 MeV excitation energy with respect to the lowest lying
$0^{-+}$. This large gap is absent in our constituent model, which
can largely be traced to the tensor force.

The $(n,\ L) = (2,\ 0)$ and (3, 0) states of $0^{++}$ and
$2^{++}$, are straightforward radial excitations of the $n = 1$
states. Since the lightest $0^{++}$ is especially light, the
excitation energy is more than 900 MeV. But for the $2^{++}$, it
seems that the radial excitation energy is only 300 MeV or so,
while going on to $n = 3$ states, the excitation energy is less
than 100 MeV. We notice that the size has also reached beyond 2
fm, hence the drop in excitation energy reflects the approach to
string breaking beyond a couple of fermis. Such behavior, however,
is not seen on the lattice, where the second $2^{++}$ state is
also almost 900 MeV higher than the lightest one, which again
appears more ``stringy" (``Regge").

The $(n,\ L,\ S) = (3,\ 2,\ 0)$ and (3, 2, 2) states in the
constituent model are close to degenerate, which arise again due
to a balance between spin-orbit and tensor forces. Interestingly,
the $0^{++}$ constructed out of $L = S = 2$, possible only
starting with $n = 3$, is slightly lighter than the radial
excitation of the lightest $0^{++}$. It is thus actually the
second lightest such state, with size of 1.1 fm, comparable to the
lightest $2^{++}$ but only half that of the radially excited
$0^{++}$. We note that its mass of 2685 MeV is in excellent
agreement with the second $0^{++}$ state from lattice, although
our radial excitation states are also not in disagreement. The
$(n,\ L,\ S) = (3,\ 2,\ 0)$ $2^{++}$ state is slightly lighter and
smaller in size as well than the radially excited $2^{++}$. The
interpretation of excitation energy for $0^{++}$ versus $2^{++}$
therefore could be rather different between our model and lattice
calculations.

The above two special $L = 2$ states are accompanied by the host
of $1^{++}$, $2^{++}$, $3^{++}$ and $4^{++}$ glueballs which are
comparable in mass and gradually growing in size. They could be
more interesting than the ``usual" radial excitations of the
lightest $0^{++}$ and $2^{++}$, or $0^{-+}$, $1^{-+}$ and
$2^{-+}$, which are two and a half fermi in size and rather large.
In contrast, the lattice $3^{++}$ and $4^{++}$ states are close to
3700 MeV, again appearing as stringy excitations.

The experimental results are rather uncertain at present, but
there does seem to be many states in the 2000 MeV region as
compared to lattice suggestions. In part for reasons of
comparison, we take here 1980 MeV for a $2^{++}$ candidate
mass~\cite{Bugg}. Glueball candidates abound but they are very
hard to pin down. In contrast to lattice results, especially for
the glueballs with size less than 2 fm, the richness of our
spectrum should give some hope for experimental search, although
one clearly has a long way to go. For the lattice calculations, it
is not clear to us whether the short distance spin-orbit and
tensor interactions, of great importance to our model (though
mysteriously balancing each other), are replicated.

In our model, the size of the lightest scalar glueball is of order
0.1 fm. The extreme smallness may be an artifact of the treatment
of the attractive $\delta$ potential, but we do expect a
physically smaller lightest scalar from the heuristic point of
view, because of the extra attraction. A more direct calculation
of the $0^{++}$ glueball mass and size on the lattice would
require relatively fine lattice spacings~\cite{Morningstar2}. It
would be interesting to see if our result of small $0^{++}$ size
could be born out on the lattice. The lightest pseudoscalar
glueball mass is 2570 MeV. However, there are the other two
$0^{-+}$ states with mass 2765 and 3780 MeV composed of two- and
three-gluons, respectively. Mixing among them should make the
lightest $0^{-+}$ two-gluon glueball mass lighter. On the other
hand, despite the lattice QCD prediction that the lowest lying
$1^{-+}$ glueball has mass heavier than $J/\psi$~\cite{Bali2}, one
should not exclude the $1^{-+}$ glueball in search below 3
GeV~\cite{Li3}. The $1^{--}$ and $3^{--}$ only exist in
three-gluon glueball states, hence their masses should be heavier
than 3 GeV. Scaling from the lattice result of
$M_{2^{++}}$=2000--2400 MeV \cite{Morningstar,Weingarten,Liu} the
mass range of these glueballs is 3.1-3.7 GeV, right in the
ballpark of $J/\psi$ and $\psi'$ masses. The proximity of the
$1^{--}$ glueball to $J/\psi$ or perhaps $\psi'$ may be what is
needed from comparison of $J/\psi$ and $\psi'$ two body hadronic
decays~\cite{Suzuki}.

In conclusion, two-gluon glueballs have been studied in a
potential model with constituent gluons. The potential is
approximated by one-gluon-exchange plus a breakable string. The
mass and size of radial and orbital excitations, up to principle
quantum number $n=3$, are evaluated by variational method. With
the lowest lying $0^{++}$ and $2^{++}$ masses from lattice taken
as input, all masses of two-gluon glueballs are found to be below
3 GeV with size less than 3 fm. The predicted masses for the
second $0^{++}$ as well as the lowest $0^{-+}$ are in excellent
agreement with lattice. Further excitation patterns, however,
differ considerably. While lattice calculations find fewer states
populating 3000 MeV range, hence an excitation energy of 500--900
MeV, our constituent model gives two more $0^{++}$ states, four
more $2^{++}$ states, and an additional $0^{-+}$ state, which are
12\%-17\% heavier than the lightest tensor. This pattern may well
be more consistent with experiment. There are also a number of
$1^{\mp+}$, $2^{-+}$, $3^{++}$ and $4^{++}$ glueballs. Besides a
rather small size for the lightest scalar, there is one $0^{++}$,
$2^{++}$, $0^{-+}$, $1^{-+}$ and $2^{-+}$ each that are 1 fm in
size, each one the lightest member for the given quantum number,
with $1^{++}$ only slightly larger. The number of glueballs
clustering at 13\% or more heavier than the lightest $2^{++}$ are
typically 2 fm in size. Mass and size seem positively correlated.
While the model certainly has its limitations, its heuristic
nature may provide some help to the long quest for uncovering
glueball states in Nature.

\vskip 0.3cm \noindent{\bf Acknowledgement}.\ \ This work is
supported in part by the National Science Council of R.O.C. under
Grant NSC-90-2112-M-002-022.

%%%%%%%%%%%%%%%%%%%%%%%%%%%%%%%%%%%%%%%%%%


\begin{references}
\bibitem{foot}
In general, $\lambda$-dependence can be neglected,
and the larger the $\beta$, the heavier the glueball.
\end{references}
\end{document}